\documentclass[final]{svjour2}
\usepackage{graphicx}
\usepackage{rotating}
\usepackage{amssymb}
\usepackage{mathptmx}
\usepackage[numbers]{natbib}
\makeatletter
\journalname{Journal of Low Temperature Physics}

\bibpunct{}{}{,}{s}{}{,}

\begin{document}

\newcommand{\hdblarrow}{H\makebox[0.9ex][l]{$\downdownarrows$}-}
\title{Large-area Reflective Infrared Filters for Millimeter/sub-mm Telescopes}

\author{Z. Ahmed \and J.A. Grayson \and K.L. Thompson \and C-L. Kuo \and G. Brooks \and T. Pothoven}

\institute{Z. Ahmed \and J.A. Grayson \and K.L. Thompson \and C-L. Kuo \at Department of Physics, Stanford University,\\ Stanford, CA 94306, USA\\
\email{zeesh@stanford.edu}
\and K.L. Thompson \and C-L. Kuo
\at Kavli Institute for Particle Astrophysics and Cosmology,\\ Stanford, CA 94305, USA
\and G. Brooks \and T. Pothoven
\at Laserod Technologies LLC,\\ Torrance, CA 90501, USA
}

\date{XX.XX.20XX}

\maketitle

\begin{abstract}
Ground-based millimeter and sub-millimeter telescopes are attempting to image the sky with ever-larger cryogenically-cooled bolometer arrays, but face challenges in mitigating the infrared loading accompanying large apertures. Absorptive infrared filters supported by mechanical coolers scale insufficiently with aperture size. Reflective metal-mesh filters placed behind the telescope window provide a scalable solution in principle, but have been limited by photolithography constraints to diameters under 300\,mm. We present laser etching as an alternate technique to photolithography for fabrication of large-area reflective filters, and show results from lab tests of 500-mm-diameter filters. Filters with up to 700-mm diameter can be fabricated using laser etching with existing capability.
\end{abstract}

\section{Introduction}
Telescopes observing the millimeter and sub-millimeter sky are increasingly employing bolometer arrays operated at sub-kelvin temperatures to achieve detector noise at or near the photon-shot noise limit. In this scenario, sensitivity scales with the square-root of number of bolometer pixels on the sky. Substantial increases in sensitivity for future telescopes such as BICEP3 \cite{kuo}, POLAR-1 \cite{kuo}, Polarbear-2 \cite{polarbear} and SPT-3G \cite{spt3g} will rely on much larger focal planes accompanied by apertures ranging from 25\% to more than 100\% larger in size than the typical $\sim300$\,mm in current telescopes. An increase in aperture size will require a corresponding increase in cooling power to the bolometer array or a mitigation of infrared loading on the system from the telescope window and sky. Closed-cycle cryocoolers are replacing expensive open-cycle liquid cryogens, leading to cooling capacity limitations of $O(30)$\,W at 40\,K and $O(1)$\,W at 4\,K. This cooling power is typically coupled to plastic filters that absorb infrared radiation while being transparent to signal. Any infrared power greater than the capacity of the cryocooler must be reflected away before entering the cold optics of the telescope. This is achieved by metal-mesh reflective filters tuned for maximum transmission in the signal band and reflection of infrared power.  

\begin{figure}
\begin{center}
\includegraphics[%
  width=0.4\linewidth,
  keepaspectratio]{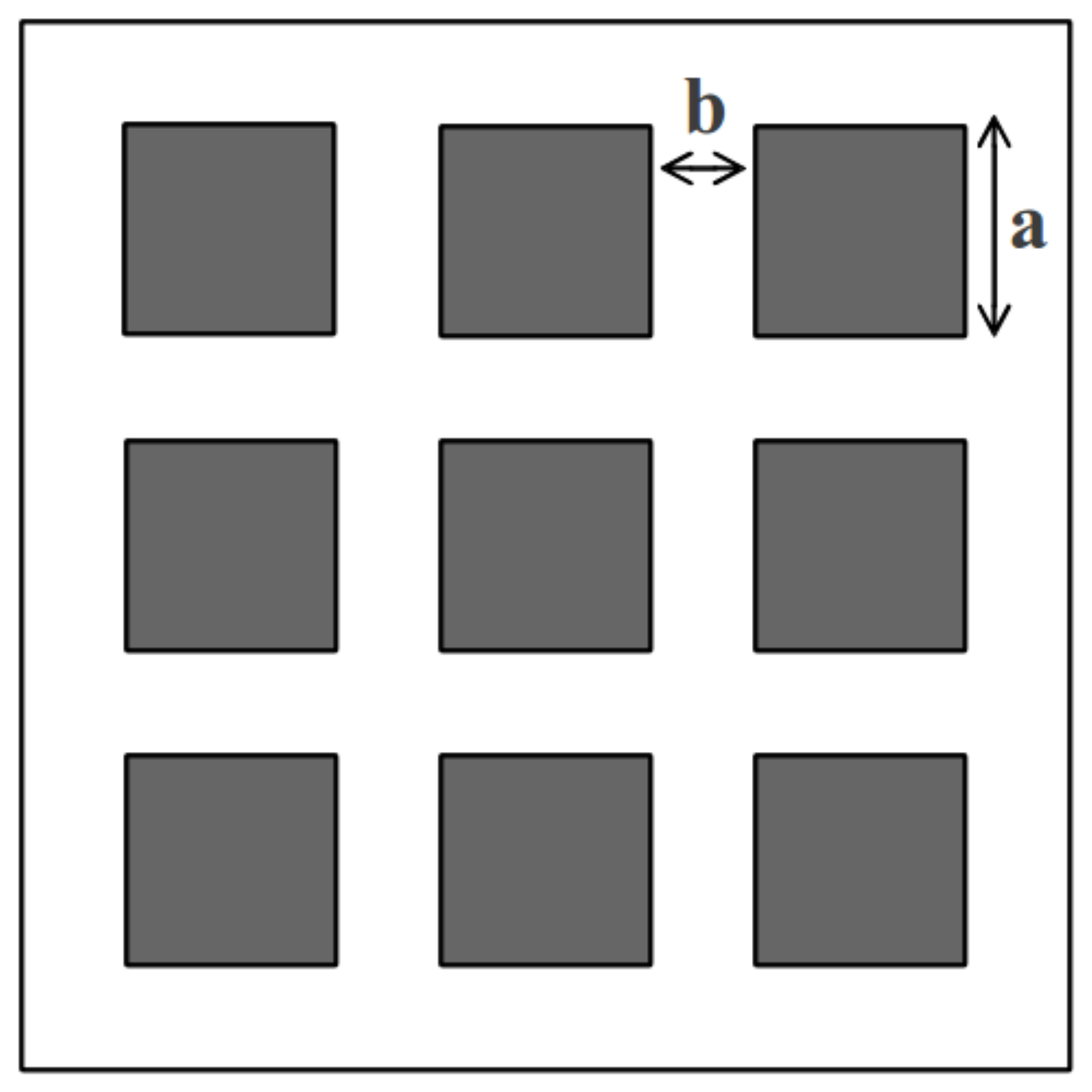}
  \end{center}
\caption{Schematic of low-pass filter consisting of metal squares on dielectric. Here, 'a' denotes the size of the square side and 'b' denotes the edge-to-edge space between squares as a reference for Figure \ref{plot} }
\label{schematic}
\end{figure}

\begin{figure}
\begin{center}  
  \includegraphics[%
  width=.8\linewidth,
  keepaspectratio]{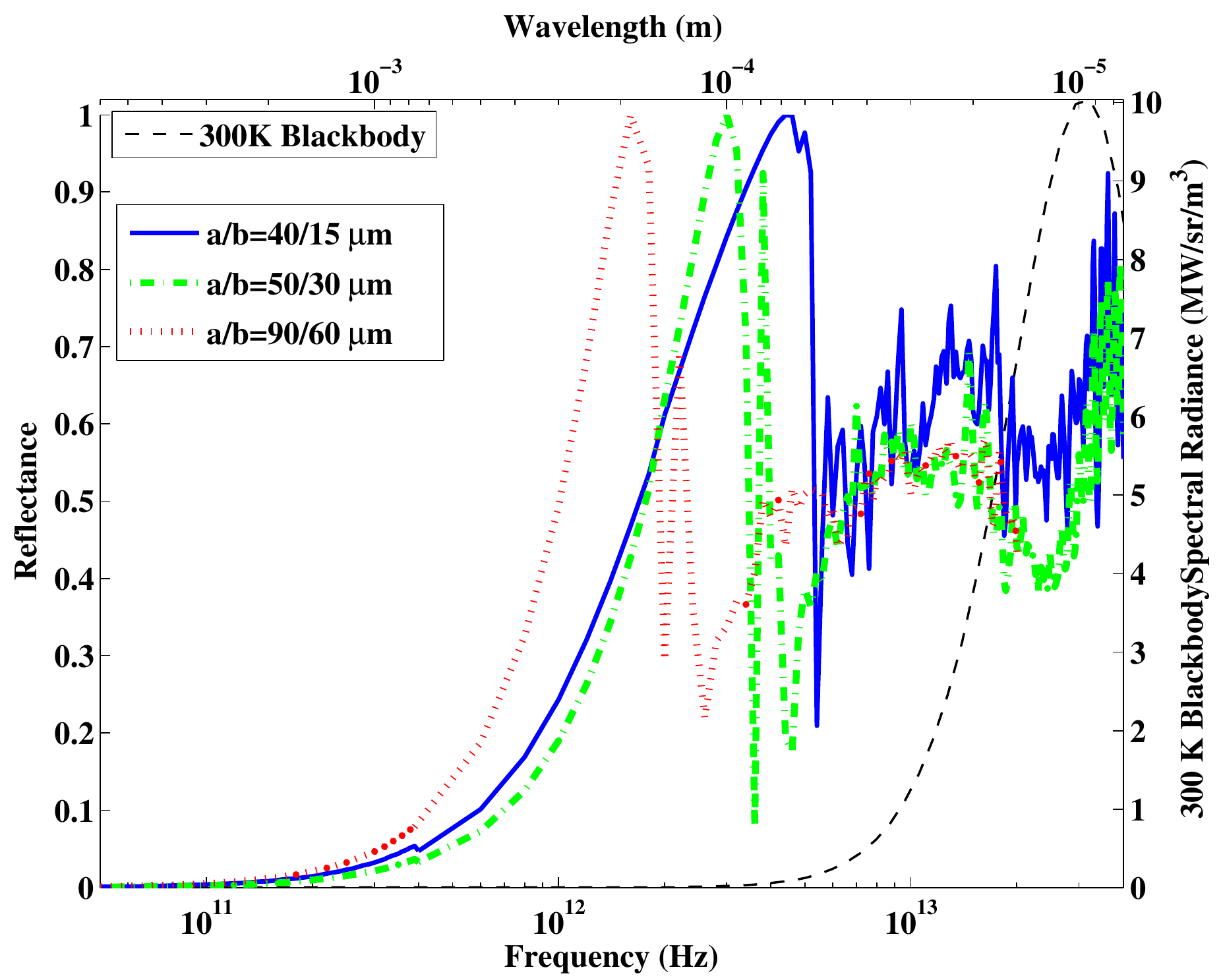}
\end{center}
\caption{(Color online) Reflectance of three different capacitive low-pass patterns as a function of frequency, simulated in HFSS. In each case, 'a' denotes the size of the square side and 'b' denotes the space between squares, edge-to-edge. Refer to Figure \ref{schematic} for schematic of metal pattern. Also plotted is the spectral radiance of a 300K blackbody.}
\label{plot}
\end{figure}

Metal-mesh filters and their characteristics have been described well by Ulrich \cite{ulrich}. The latest production techniques have been developed and documented by Ade et al \cite{tucker, ade}. For the purposes of these proceedings, we remind the reader that a repeating grid of thin metal squares or metal lines or both, suspended on a thin dielectric such as Mylar or polypropylene will act as a low-pass or high-pass or band-defining filter, respectively.  The metal features' sizes are approximated by modeling an effective capacitive and/or inductive lumped element model in the free space transmission of the electromagnetic wave, to obtain desired filtering properties.  
Simulations provide the exact spectral response for a filter, taking into account diffraction and substrate absorption characteristics.  As an example, we used High Frequency Structure Simulator (HFSS) to simulate a simple capacitive mesh filter. The model consisted of a metallic square (perfect conductor) on a 3.5\,$\mu$m-thick dielectric substrate (Mylar), with master/slave (periodic) boundary conditions on four sides.  A schematic of the intended pattern is shown in Figure \ref{schematic}. Electromagnetic excitation is provided as a plane wave source 100\,$\mu$m above the substrate and aimed at the pattern. In Figure \ref{plot}, we plot the the integrated far-field reflected power normalized to incident power as a function of frequency for three different patterns. We observe that features sizes of $O(10)\,\mu$m produce near-perfect acceptance of millimeter waves, with cut off frequencies in the terahertz region and reflection of a significant fraction of infrared power.  Stacks of such simple capacitive patterns act as thermal filters in cryogenic telescopes to broadly reduce infrared loading, which peaks at $O(10)\,\mu$m wavelengths for 300K blackbody radiation. For more precise edge- or band-defining filters, multiple layers with slightly different spectral characteristics can be hot pressed or stacked with precision spacing to achieve sharper cut-offs and cancel resonant features \cite{ade}.  Individual layers of the thermal or band-defining filter stacks are traditionally fabricated using standard photolithography techniques, which practically limit the diameters of these filters to no greater than 300\,mm.

Millimeter and sub-millimeter telescopes in the coming years will benefit from large apertures and increased throughput, requiring filters with diameters $\sim400-700$\,mm.  Metal-mesh filters for such telescopes will require a doubling of currently achievable filter dimensions with high precision metal features over large areas. This is possible at great expense by increasing the size of the photolithographic evaporation chamber and employing a custom-built UV system.  In these proceedings, we present and demonstrate an alternate cost-effective technique,  replacing photolithography by laser etching in the fabrication process for metal-mesh filters.

\section{Laser ablation of metal on dielectric}
Instead of attempting to directly pattern metal features over large areas of dielectric, we chose to first deposit metal on dielectric in a layer of uniform thickness and then use laser ablation to remove unwanted features. This simplified the process of patterning into two steps, both of which individually provide sufficient precision over areas much larger than those accommodated in standard laboratory photolithography setups.

\begin{figure}
\begin{center}
\includegraphics[%
  width=0.65\linewidth,
  keepaspectratio]{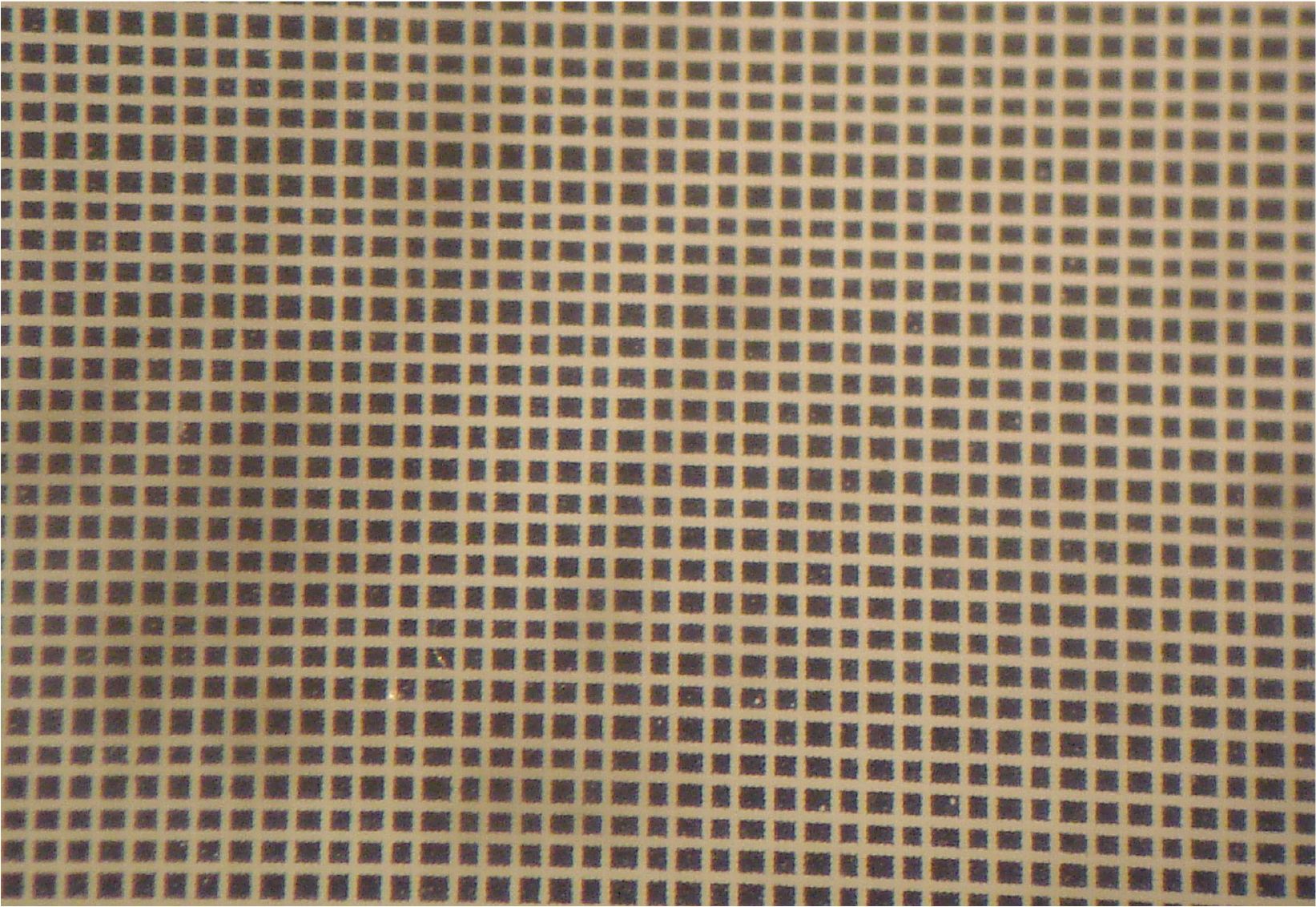}
  \end{center}
\caption{(Color online) Microscope photograph of prototype
   filter produced by ablating aluminum on Mylar with a
   galvanometer-mounted ultraviolet laser, achieving 30\,$\mu$mm
   features.}
\label{pattern}
\end{figure}

For the first prototype filters, we used Mylar for the dielectric material and aluminum for metallization. This choice was dictated purely by ease of procurement, with the knowledge that other combinations such as copper on polypropylene were equally viable. We sourced 3.5\,$\mu$m-thick Mylar from ER Audio\footnote{546 Brookton Highway, Roleystone 6111, Western Australia, Tel: +61-8-93976212, Email: orders@eraudio.com.au} and used Deposition Sciences\footnote{3300 Coffey Lane 
Santa Rosa, CA 95403, Email:solutions@depsci.com, Tel:+1-707-573-6700} to vapor deposit 0.04\,$\mu$m-thick  aluminum. We mounted the aluminized Mylar in a 500-mm diameter jig, designed to fit directly into a test cryostat. Then, at Laserod Technologies LLC\footnote{20312 Gramercy Pl.
Torrance, CA 90501, Email: info@laserod.com, Tel: +1-310-328-5869}, we used a frequency-tripled Nd:YAG laser with 355\,nm wavelength, directed by a mirror galvanometer, to etch channels in the metal layer. The laser power and write speed to ablate 0.04 $\mu$m-thick aluminum were determined empirically by testing on a small reference sample of aluminized Mylar. The smallest laser spot size achieved on the system was 30\,$\mu$m. Thus, the prototype filters were produced with 50\,$\mu$m metal squares separated by 30\,$\mu$m gaps. A picture of the pattern achieved for the prototype is shown in Figure \ref{pattern}.  Note that instability in the galvanometer introduced $\sim10\,\mu$m periodic variation in the laser spot positioning. This has since been mitigated by using a precision XY-traveling stage at the expense of increased write time.

\section{Performance of prototype filters}
\begin{figure}
\begin{center}  
  \includegraphics[%
  width=0.65\linewidth,
  keepaspectratio]{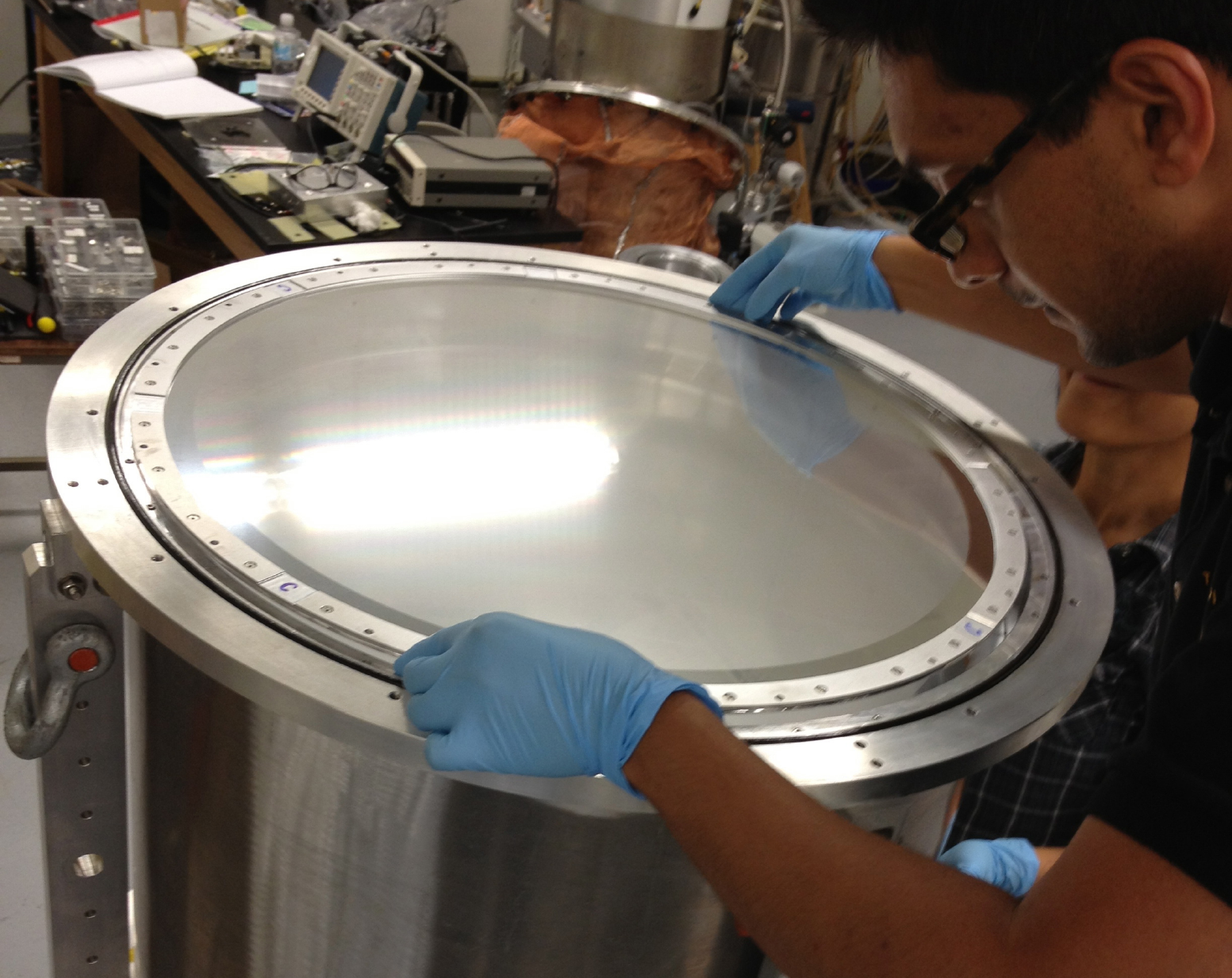}
\end{center}
\caption{(Color online) Installation of prototype 500-mm metal-mesh filters in test cryostat.}
\label{installation}
\end{figure}

 We tested the prototype filters for general infrared and microwave transmission properties. For the infrared measurement, a single filter was placed between a source and a thermopile detector sensitive to wavelengths 0.6--40$\,\mu$m (7.5--500\,THz in frequency). The source was modulated between a 77\,K liquid nitrogen load and a 300\,K blackbody. For the microwave measurement, we used a 150\,GHz noise source and a diode detector, both coupled to free space via feed horns. The infrared transmission of a single filter was measured to be $35\pm1\%$ and the microwave transmission to be $98\pm1\%$.  A stack of six such filters was installed behind the window of a 560-mm window diameter test cryostat (Figure \ref{installation}). The filters were oriented plane parallel to each other, but with random angular alignment of the metal patterns and with $6\pm1$\,mm spacing between filters.  The cryostat was expected to receive $\approx 115\,$W of loading from infrared radiation of ambient 300\,K. Cooling was provided by a Cryomech PT-415 mechanical cooler, with its first stage coupled to a stack of absorptive plastic filters (PTFE 12.7\,mm, PTFE 34.3\,mm, Nylon 5.3\,mm thicknesses) to absorb unreflected infrared radiation. With the mesh filters in place, the measured load on the first stage was 19\,W, well within the 36\,W cooling capacity of the cooler. This demonstrated the viability of laser-ablation as a technique to produce working large-area capacitive mesh thermal filters. 
 
\begin{figure}
\begin{center}
\includegraphics[%
  width=0.55\linewidth,
  keepaspectratio]{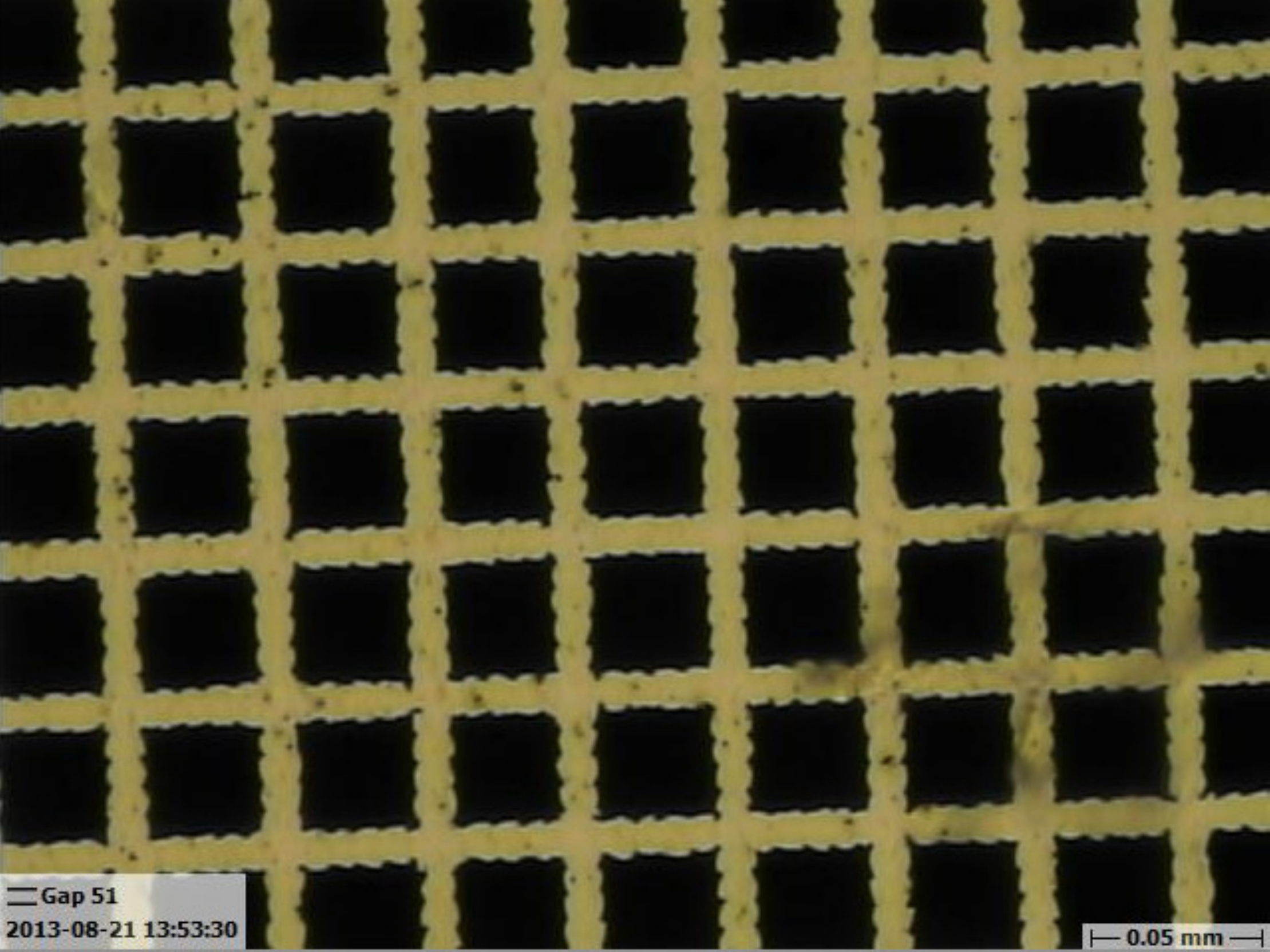}
  \end{center}
\caption{(Color online) Microscope photograph of a capacitive mesh 
   filter produced by ablating aluminum on Mylar with an XY-stage-mounted
    ultraviolet laser, achieving 15\,$\mu$m features.}
\label{xypattern}
\end{figure}

\section{Conclusions}
The single filter fabrication technique demonstrated here is promising, even without much optimization in the first attempt. Note that during our first tests of these filters, we did not attempt to measure the thermal conductance, emissivity or operating temperatures of the prototype filters to quantify the direct thermal loading from the filters. We will measure this in the future; optimization of the dielectric and metal materials and thicknesses will enable improvements if needed. Additionally, measuring the spectral response of the filters will provide feedback for fine-tuning of the metal feature sizes to further improve performance. Since the LTD-15 conference, the ultraviolet laser used for patterning the filters was mounted in a high-precision XY-traveling stage, making possible laser spot sizes to 15\,$\mu$m with travel of up to 700\,mm. This reduced systematic errors in laser spot positioning and has enabled production of filters with smaller metal features. This in turn has led to higher frequency cut offs and thus better infrared rejection as suggested in Figure \ref{plot}. A sample pattern with 40\,$\mu$m squares and 15\,$\mu$m channels, printed on a 640-mm diameter filter is shown in Figure \ref{xypattern}. In the future, the use of laser ablation in patterning single metal-mesh filters can be incorporated with traditional filter fabrication techniques to produce high-performance air gap and hot-pressed filter stacks with fine-tuned band properties.  This development overcomes a substantial cost challenge in proceeding to larger throughput telescopes in  millimeter and sub-millimeter astrophysics.

\begin{acknowledgements}
This material is based upon work supported by the National Science Foundation (NSF) under Grant No. 1056465. J. A. Grayson is supported by an NSF Graduate Research Fellowship under Grant No. DGE-114747.
\end{acknowledgements}



\begin{thebibliography}{99}


\bibitem{kuo}
C-L. Kuo, {\it Proc. of the Int. Astron. Union} \textbf{8}, 80, (2012).

\bibitem{polarbear}
T. Tomaru et al, {\it Proc. SPIE 8452, Millim. and Submillim. Detect. and Instrum. for Astron. III} \textbf{8452}, (2012).

\bibitem{spt3g}
B. Benson, {\it Proc. of the Int. Astron. Union} \textbf{8}, 76, (2012).

\bibitem{ulrich}
R. Ulrich, {\it Infrared Phys.} \textbf{7}, 37, (1967).

\bibitem{tucker}
C.E. Tucker, P.A.R. Ade, {\it Proc. SPIE 6275, Millim. and Submillim. Detect. and Instrum. for Astron. III} \textbf{6275}, (2006).

\bibitem{ade}
P.A.R. Ade, G. Pisano, C. Tucker, S. Weaver, {\it Proc. SPIE 6275, Millim. and Submillim. Detect. and Instrum. for Astron. III} \textbf{6275}, (2006)

\end{thebibliography}
\end{document}